\newcommand{\myname}{Geoff Boeing, Clemens Pilgram, and Yougeng Lu}
\newcommand{\myemail}{boeing@usc.edu}

\newcommand{\papertitle}{Urban Street Network Design and Transport-Related Greenhouse Gas Emissions around the World}
\newcommand{\papercitation}{Boeing, G., C. Pilgram, and Y. Lu. 2024. \papertitle. \textit{Transportation Research Part D: Transport and Environment}, 127, 103961. \href{https://doi.org/10.1016/j.trd.2023.103961}{doi:10.1016/j.trd.2023.103961}}
\newcommand{\paperkeywords}{air pollution, climate change, street networks, transport planning, urban design, urban planning}

\RequirePackage[l2tabu,orthodox]{nag}     % warn if using any obsolete or outdated commands
\documentclass[12pt,letterpaper]{article} % document style

\usepackage[T1]{fontenc}    % output T1 font encoding (8-bit) so accented characters are a single glyph
\usepackage[utf8]{inputenc} % allow input of utf-8 encoded characters
\usepackage{ebgaramond}     % document's serif font
\usepackage{tgheros}        % document's sans serif font

\usepackage[strict,autostyle]{csquotes} % smart and nestable quote marks
\usepackage[USenglish]{babel}           % automatically regionalize hyphens, quote marks, etc
\usepackage{microtype}                  % improves text appearance with kerning, etc

\usepackage{abstract}                   % allow full-page title/abstract in twocolumn mode
\usepackage{amsmath}
\usepackage{array}
\usepackage{authblk}                    % footnote-style author/affiliation info
\usepackage{booktabs}                   % better looking tables
\usepackage{caption}                    % custom figure/table caption styles
\usepackage[final]{draftwatermark}      % watermark paper as a draft
\usepackage{endnotes}                   % enable endnotes
\usepackage{geometry}                   % configure page dimensions and margins
\usepackage{graphicx}                   % better inclusion of graphics
\usepackage{hyperref}                   % hypertext in document
\usepackage{longtable}
\usepackage{multirow}
\usepackage[section]{placeins}
\usepackage{natbib}                     % author-year citations w/ bibtex, including textual and parenthetical
\usepackage{pdflscape}                  % rotate wide tables or figures on a page to make them landscape
\usepackage{setspace}                   % configure spacing between lines
\usepackage{titlesec}                   % custom section and subsection heading
\usepackage{url}                        % make nice line-breakble urls
\usepackage{threeparttable}
\usepackage{ragged2e}                   % more table formatting options
\usepackage{tabularx}                   % yet more table formatting options
\usepackage{subcaption}

\graphicspath{{./figures/}}

\titleformat{\section}{\normalfont\sffamily\large\bfseries\color{black}}{\thesection.}{0.3em}{}
\titleformat{\subsection}{\normalfont\sffamily\small\bfseries\color{black}}{\thesubsection.}{0.3em}{}

\hypersetup{
	pdfauthor={\myname},
	pdftitle={\papertitle},
	pdfsubject={\papertitle},
	pdfkeywords={\paperkeywords},
	pdffitwindow=true,         % window fit to page when opened
	breaklinks=true,           % break links that overflow horizontally
	colorlinks=false,          % remove link color
	pdfborder={0 0 0}          % remove link border
}
\begin{document}

\title{\papertitle\footnote{Citation info: \papercitation~Correspondence: \href{mailto:\myemail}{\myemail}}}
\author[]{\myname}
\affil[]{}%\myaffiliation}
\date{}

\maketitle

\begin{abstract}
This study estimates the relationships between street network characteristics and transport-sector CO\textsubscript{2} emissions across every urban area in the world and investigates whether they are the same across development levels and urban design paradigms. The prior literature has estimated relationships between street network design and transport emissions---including greenhouse gases implicated in climate change---primarily through case studies focusing on certain world regions or relatively small samples of cities, complicating generalizability and applicability for evidence-informed practice. Our worldwide study finds that straighter, more-connected, and less-overbuilt street networks are associated with lower transport emissions, all else equal. Importantly, these relationships vary across development levels and design paradigms---yet most prior literature reports findings from urban areas that are outliers by global standards. Planners need a better empirical base for evidence-informed practice in under-studied regions, particularly the rapidly urbanizing Global South.
\end{abstract}

\section{Introduction}

Earth's climate is changing, characterized by rising temperatures and increasingly erratic weather events such as flooding, drought, landslides, and hurricanes \citep{doi:10.1126/science.1150195}. The growing frequency and magnitude of these phenomena adversely affect human life \citep{hess-19-2247-2015, Myhre2019}. Greenhouse gas (GHG) emissions constitute one of the principal drivers of climate change, and transport accounts for a substantial share of such emissions \citep{Lashof}. For example, roughly one-third of US GHG emissions arise from transport \citep{us_epa_2021}, as does 29\% of emissions in Europe \citep{eeg_2018}. Around the world, rapid urbanization and ever-growing automobile dependence have led to massive growth in transport demand and distances traveled---and, in turn, transport-sector GHG emissions.

Sustainable urbanization requires planners to better understand the climate impacts of their plans. To this end, many recent studies have sought to measure the relationship between urban form---particularly street network design---and transport emissions. Variables such as street length, straightness, and intersection density appear in many such studies \citep{HANDY200264, HANKEY20104880, HONG201447, MOHAJERI2015116, doi:10.1080/00420980601136588}. However, most existing studies focus on individual countries or regions, particularly those with plentiful high-quality data, and usually collect relevant data from local or regional organizations \citep{boeing_using_2022}. A lack of consistent global data has stymied investigating relationships worldwide \citep{boeing_street_2021}. This poses problems for both scientific generalizability and evidence-informed practice \citep{giles-corti_creating_2022}. Better and more-comprehensive local estimates of the relationship between street network design and transport emissions would provide an important evidence base for planners to build more sustainable cities.

This study addresses this need and asks: what is the relationship between street network design and transport-sector CO\textsubscript{2} emissions across all of the world's urban areas---and to what extent is this relationship heterogeneous between different kinds of cities? Using street network characteristics and data on  transport-sector emissions, we estimate relationships between street network design and corresponding urban area transport CO\textsubscript{2} emissions, including how these relationships vary across different development levels and urban design paradigms. Using a global ordinary least squares (OLS) regression model with a full set of controls, we find significant \textit{ceteris paribus} associations between several street network design variables and  transport-sector CO\textsubscript{2} emissions. Broadly consistent with the prior smaller-sample and regional literature \citep[e.g.,][]{Boarnet, doi:10.1080/01944363.2011.593483, doi:10.1177/08854120022092890, doi:10.1080/01944361003766766, HANKEY20104880}, our worldwide results reveal that straighter, more-connected, and less-overbuilt street networks are associated with lower transport CO\textsubscript{2} emissions. However, important differences exist between different types of urban areas, which we unpack through spatial regimes models of UN development groups and design paradigms discovered through a cluster analysis.

This study is the first to comprehensively estimate the relationships between street network design and transport emissions across all of the world's urban areas. It offers practitioners both generalized estimates of these relationships for evidence-informed sustainability planning and paradigm-specific estimates for better local applicability. More specifically, it allows insights into how useful studies from other parts of the world may be to settings other than where they were conducted. In doing so, this study also offers researchers a new set of data and methods to conduct further case studies to identify localized relationships through a novel method of clustering cities based on street network attributes.

The rest of this article is organized as follows. First we briefly review the literature of street network design and transport emissions, highlighting its traditional limitations. Next we describe our methods for measuring street network design, transport emissions, and their relationships worldwide, as well as how we cluster urban areas. Then we present our findings on these global and heterogeneous relationships. Finally we discuss their implications for both transport research and practice around the world before concluding.

\section{Background}

An extensive literature explores and tests how the built environment influences people's travel behavior, and with it, their emissions. Such studies have converged on a set of central factors---such as density, diversity, and design, the \enquote{three Ds} originally identified by \citet{CERVERO1997199}---as dimensions of the built environment informing variable selection and model specification.

Typically, studies of how urban form relates to emissions rely on case study research utilizing data from one site \citep[e.g.,][]{doi:10.1080/00420980601136588,HONG201447,CAO2017480,XU20181336} or from several cities within the same country \citep[e.g.,][]{MOHAJERI2015116,HANKEY20104880,doi:10.1080/01944363.2010.486623, WANG2017189}. Such studies have frequently investigated Europe \citep[e.g.,][]{MOHAJERI2015116, doi:10.1080/00420980601136588}, North America \citep[e.g.,][]{doi:10.1061/(ASCE)0733-9488(1984)110:1(1), HONG201447, HANKEY20104880, doi:10.1080/01944363.2010.486623}, and China \citep[e.g.,][]{CAO2017480, XU20181336, WANG2017189}. However, other world regions---such as South and Southeast Asia, Latin America, and Africa---remain underrepresented. Research designs also vary widely, comprising both travel surveys---where each observation represents an individual's travel and associated emissions \citep{HONG201447}---and aggregate-level simulations---where each observation represents an entire city, urban area, or other areal unit \citep{HANKEY20104880, WANG2017189, MOHAJERI2015116}.

While measuring density in some form and including it as an explanatory variable is common across almost all studies, operationalizations of density---and what other variables are included---vary dramatically. Emissions themselves can be measured as fuel consumption \citep{doi:10.1061/(ASCE)0733-9488(1984)110:1(1),Siew_Yin_Chin_Siong_2010}, carbon emissions \citep{MOHAJERI2015116,CAO2017480}, or as a variety of other pollutants of interest such as ozone \citep{doi:10.1080/01944363.2010.486623} or particulate matter \citep{doi:10.1177/00420980221145403}. Population density---that is, the number of residents in an urban area divided by its area---is almost always included as a variable, but land use mix and other built environment dimensions like accessibility are less common, particularly when using aggregate units of observation \citep{doi:10.1080/01944361003766766,HONG201447}. Measures of street network design also vary. Researchers define network density variously as intersection density \citep[e.g.,][]{HONG201447}, edge density \citep[e.g.,][]{WANG2017189}, street length \citep{MOHAJERI2015116}, or the area occupied by street networks \citep{MOHAJERI2015116}. Similarly, control variables included in studies often reflect local data availability or unique sociodemographic variables relevant to the particular study area, including such things as industrial composition \citep{WANG2017189} or a person's migrant status \citep{HONG201447} in Chinese studies, or income levels and vehicle ownership in Europe \citep{doi:10.1080/00420980601136588}.

In sum, the heterogeneity in research designs and model specifications makes it challenging to compare these studies directly, generalize universal theory from their diverse estimates, or apply their evidence in practice in different urban contexts. Nevertheless, these studies' findings converge on certain principles: greater intersection density, street connectedness, land-use diversity, and pedestrian-oriented design are associated with less automobile travel, in turn potentially leading to lower transport emissions \citep{Boarnet, doi:10.1080/01944363.2011.593483, doi:10.1177/08854120022092890, doi:10.1080/01944361003766766, HANKEY20104880}. Regardless of how exactly these variables are defined, greater population densities and shorter streets are consistently associated with lower transport emissions \citep{MOHAJERI2015116, doi:10.1080/00420980601136588, HONG201447, XU20181336, WANG2017189} and lower neighborhood exposure to pollutants \citep{doi:10.1080/01944363.2010.486623}.

Planners and policymakers need to understand the interconnectedness of urban form and transport emissions on a global scale. This becomes particularly crucial for benchmarking and monitoring cities with comparable urban characteristics. While existing research predominantly examines land use and built-up area to assess the physical structure of cities \citep{LEMOINERODRIGUEZ2020103949, uhl2021century}, some studies also incorporate socioeconomic factors like population size and density as indicators of urban form \citep{frenkel2008measuring}. Recent studies have also used morphological configurations of urban landscapes \citep{TAUBENBOCK2020102814} and polycentricity for city clustering \citep{JUNG2022101223}.

However, urban research and practice are context-specific. While most of this literature focuses on the Global North and large cities---often due to data availability and researcher familiarity---many of the most pressing urban and environmental challenges exist in the Global South and in smaller cities \citep{McPhearson2016}. Different development trajectories, social contexts, and geographies mean that coefficient estimates and lessons learned in, for instance, the United States are not necessarily applicable to cities in less-developed countries. The literature's geographical contexts and heterogeneity in specifications pose challenges for generalizability and applying findings to places unlike those originally studied. To better serve practitioners, planning scholars must expand research in less-studied and rapidly changing urban contexts \citep{giles-corti_what_2022}.

\section{Methods}

This study takes up this challenge to estimate these relationships worldwide while also unpacking geographical heterogeneity. It estimates the \textit{ceteris paribus} relationships between urban area street network design and transport emissions, then tests how these relationships differ between development contexts. Next it uses a cluster analysis to discover street network design paradigms and then retest these relationships between design paradigms. Our data measure the same variables the same way for all the places we study.

\subsection{Data}

We use the European Union's Global Human Settlement Layer's Urban Centres Database (UCD), which reports data on every urban area around the world \citep{florczyk_description_2019}. These urban areas are defined, per the United Nations Statistical Commission's methodology, by having a population of at least 50,000 across contiguous 1×1 kilometer grid cells with a density of at least 1,500 inhabitants per square kilometer based on satellite data, censuses, and other inputs \citep{dijkstra_how_2020}. Table \ref{tab:city_sizes} lists our urban area counts by world region, as well as population summary statistics, after excluding urban areas missing emissions data or with fewer than 100 street network nodes.

Table \ref{tab:variable_descriptions} lists the variables used in this study alongside their summary descriptions, units, and sources. To measure street network design, we use the OSMnx package to model street network characteristics from OpenStreetMap \citep{BOEING2017126}. These include the average node degree (the number of streets connected to each intersection), the average straightness of streets, the median street grade, the length of streets per capita, and the intersection density within the built-up area, for each urban area in the UCD \citep{boeing_street_2021}. Many researchers have validated OpenStreetMap's completeness and accuracy over the years \citep{barron_comprehensive_2014,basiri_quality_2016,corcoran_analysing_2013,haklay_how_2010,zielstra_assessing_2013}. \citet{barrington-leigh_worlds_2017} caveat that Chinese OpenStreetMap data was relatively incomplete as of 2016 due to national restrictions on geospatial data. OpenStreetMap data are imperfect, but they represent the state-of-the-art today and the best available worldwide data.

\begin{table*}[tbp]
    \centering
    \footnotesize
    \caption{Urban area counts and population summary statistics, by world region.}
    \label{tab:city_sizes}
    \begin{tabular}{lrrrrr}
        \toprule
                                   & Count & Min Pop & Median Pop & Mean Pop &    Max Pop \\
        \midrule
        Africa                     & 1,382 &  50,016 &    121,050 &  304,473 & 19,734,085 \\
        Asia                       & 4,103 &  50,012 &    136,949 &  427,439 & 40,589,878 \\
        Europe                     & 1,049 &  50,053 &    106,710 &  273,569 & 14,077,364 \\
        Latin America \& Caribbean & 1,006 &  50,179 &    106,072 &  343,191 & 19,559,564 \\
        Northern America           &   372 &  50,238 &    111,423 &  464,800 & 15,950,674 \\
        Oceania                    &    41 &  51,492 &    104,070 &  376,981 &  3,745,334 \\
        \bottomrule
    \end{tabular}
\end{table*}

\begin{table*}[htbp]
	\caption{Variables' descriptions, units, and sources.}
	\label{tab:variable_descriptions}
    \footnotesize
	\begin{tabular}{m{0.25\linewidth} m{0.35\linewidth} m{0.2\linewidth} m{0.2\linewidth}}
	\toprule
	Variable & Description and Units & Source \\
	\midrule
	CO\textsubscript{2} Emissions & The emissions of transport-related CO\textsubscript{2} per capita in 2015 (tonnes/person) & GHSL UCD \\
	\midrule
	$k$ average & Average node degree (network edges [i.e., streets] per node [i.e., intersection or dead-end]) & OpenStreetMap \\
	\midrule
	Straightness & Ratio of straightline distances between nodes to network distances between nodes  & OpenStreetMap \\
	\midrule
	Intersection Density & Density in units of 10,000 intersections per square kilometer & OpenStreetMap \\
	\midrule
	Length & Mean street segment length (meters) & OpenStreetMap \\
	\midrule
	Built up area & Built-up surface area in 2015 (km\textsuperscript{2}) & GHSL UCD \\
	\midrule
	Open space & Percentage of open space in 2015 & GHSL UCD \\
	\midrule
	Population density & Residential population per built up area in 2015 (persons/km\textsuperscript{2}) & GHSL UCD \\
	\midrule
	GDP per capita & Gross domestic product per person in 2015 (USD) & GHSL UCD \\
	\midrule
	Night light & Urban night light emission (nano-watt per steradian per square centimeter) & GHSL UCD \\
	\midrule
	Grade median &  Median absolute street grade (rise over run) & OpenStreetMap \\
	\midrule
	Airport & Dummy variable: 1 = has at least one airport, 0 = otherwise & OurAirports \\
	\midrule
	Waterport & Dummy variable: 1 = has at least one waterport, 0 = otherwise & NaturalEarth \\
	\midrule
	World region & Major geographical region (e.g., Asia, Europe, etc.) & GHSL UCD \\
	\midrule
%	UN income class indicator & LIC = Low-income, LMIC = Lower-middle income, UMIC = Upper-middle income, HIC = High-income, Other = otherwise & UN \\
%	\midrule
	UN development group indicator & LDCL = less developed countries, LDC = least developed countries, MDR = most developed countries & United Nations \\
	\bottomrule
\end{tabular}
\end{table*}

The GHSL UCD reports estimated tonnes of CO\textsubscript{2} emissions produced in 2015 by the  transport-sector from non-short cycle organic fuels---more commonly referred to as \enquote{fossil fuels}---in each of the world's urban areas \citep[pp. 36-38]{florczyk_description_2019}. It also reports estimated tonnes of CO\textsubscript{2} emissions produced in 2015 from short-cycle organic fuels---that is, biofuels---for those same urban areas. Its definition of \enquote{transport-sector} encompasses all sectors coded \enquote{1A3} by the IPCC---including road, aviation, rail, shipping, and pipeline transport \citep[p. 1303]{krey_14_2014}. While there are no data specifically reporting on-road transport CO\textsubscript{2} emissions at this granular a scale for all urban areas on the planet, on-road transport CO\textsubscript{2} emissions account for approximately 72\% of all  transport-sector emissions \citep[p. 606]{sims_transport_2014}. However, this figure can vary between different urban areas, so we include controls discussed below. The reported emissions are totals: to create our dependent variable of per-capita  transport-sector CO\textsubscript{2} emissions, we divide total  transport-sector CO\textsubscript{2} emissions from both fossil and short-cycle organic fuels for each urban area by that urban area's population (also from the UCD). Theoretically, vehicle ownership rates are a key predictor of emissions, but no consistent data exist worldwide, so we use GDP per capita and nighttime light emission as proxies capturing dimensions of development and wealth.

\subsection{Model Specification}

We model the urban areas' per capita transport emissions as a function of their street network characteristics, built-up area, population density, open space, and economic controls in Model I with the form:

\begin{equation}
	\label{eq:ols_model_nointeractions}
	\log{y} = \beta_{0} + \beta_{1} X + \epsilon
\end{equation}

where $\beta_{1}$ represents the association between each predictor in the design matrix, $X$, and the response, $y$ (per capita  transport sector CO\textsubscript{2} emissions).

For a better linear fit, we take the natural logarithm of the response and several predictors. This also allows us to interpret the coefficients of logged predictors---i.e., street length per capita, built-up area size, population density, GDP per capita (adjusted for purchasing power parity), and night light emission---as percent changes in emissions associated with percent changes in each respective predictor. Coefficients for the other predictors---i.e., average node degree ($k$ average), straightness, intersection density, percent open space, and median street grade---can be interpreted as percent changes in emissions associated with a unit change in each respective predictor.

These predictors reflect both policy decisions and local geography. For example, hilly cities' street networks often must curve with the landscape to maintain a feasible grade. Yet in cities like San Francisco, planners imposed a grid over hilly terrain, and many North American suburbs consist of curving culs-de-sac despite being built on level ground. Such urban design choices often reflect tastes of particular places and eras \citep{boeing_off_2021}. Accordingly, we control for terrain with the street grade variable to remove that effect from the coefficient on straightness. Thus, in our model, straightness represents a policy variable because we look at it while holding \enquote{hilliness} constant.

Our response variable represents emissions across the  transport sector, so we include dummy variables to control for other important such emission sources---namely aviation and shipping---by identifying whether the urban area has an airport and/or water port, to limit their confounding effects. Such dummy variables are imperfect controls but the best available for isolating the relationships of interest. We also include country-level dummy variables to control for national policy differences. All regression parameter estimations are performed using OLS, weighting observations by the urban area population to not overweight small villages at the expense of large metropolises.

In a second specification (Model II), we test whether the relationships between street network design and  transport-sector emissions are the same in both developed and less-developed countries. We do this through a spatial regimes model across UN development groups plus country dummy variables. Spatial regimes essentially runs separate regressions for each group. This allows the parameter estimates to vary between least developed countries (LDC)---e.g., Bangladesh, the Democratic Republic of the Congo, Liberia, and Nepal---less developed countries excluding least (LDCL)---e.g., China, India, Iran, Nigeria, and Turkey---and most developed regions (MDR)---e.g., the United States, Japan, Russia, and the European Union's members.

\subsection{Cluster Analysis of Urban Design Paradigms}

Although UN Development Groups capture rough groupings of development level, they ignore the heterogeneity of urbanization patterns, but the relationships between urban form and emissions could vary between latent urban design paradigms. In a third regression specification (Model III), we extend our analysis by testing whether there are differences in the relationships between urban form and  transport-sector emissions across such paradigms, with cities clustered by their urban form.

To do so, we conduct a hierarchical cluster analysis using the first three principal components of six urban form variables: $k$ average, straightness, intersection density, street length per capita, percent open space, and population density. Based on the cluster analysis dendrogram, we cut its tree at six clusters to obtain well-defined groups that also conform to theory. Similar to Model II, we estimate a spatial regimes model across design paradigm clusters and control for country dummy variables.

\section{Results}

\subsection{Design Paradigm Clusters}

\begin{figure}[htbp]
    \begin{subfigure}{0.49\textwidth}
        \includegraphics[width=\linewidth]{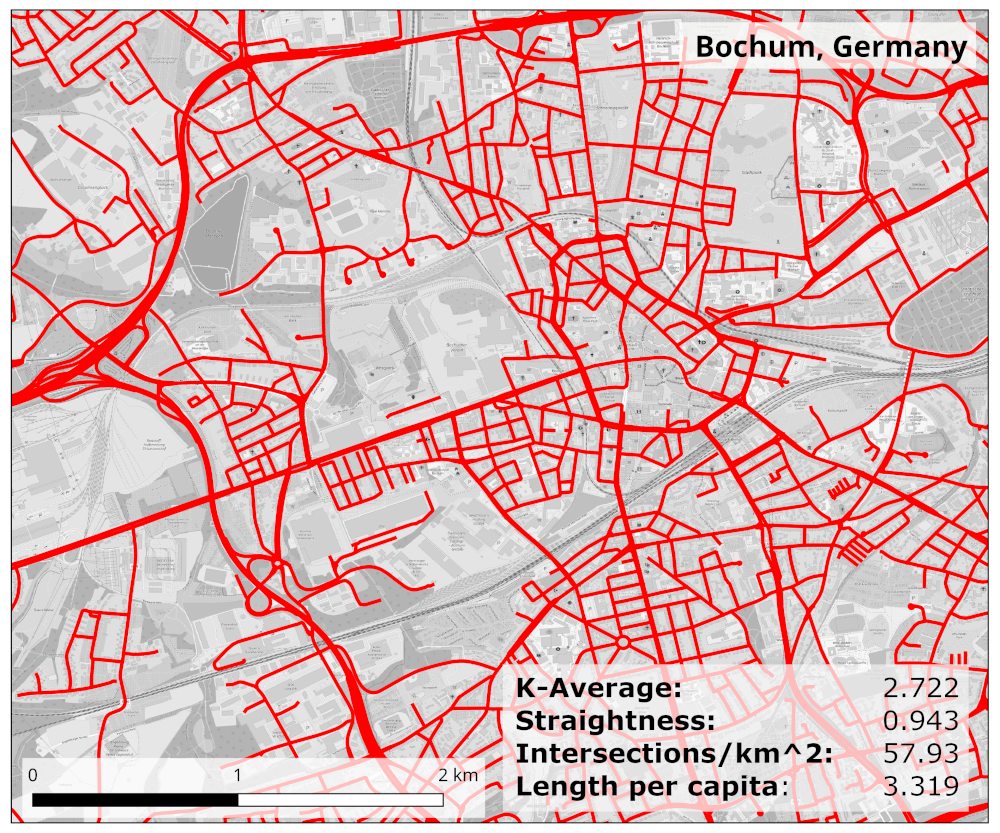}
        \caption{Circuitous and T-Intersections} \label{fig:a}
    \end{subfigure}\hspace*{\fill}
    \begin{subfigure}{0.49\textwidth}
        \includegraphics[width=\linewidth]{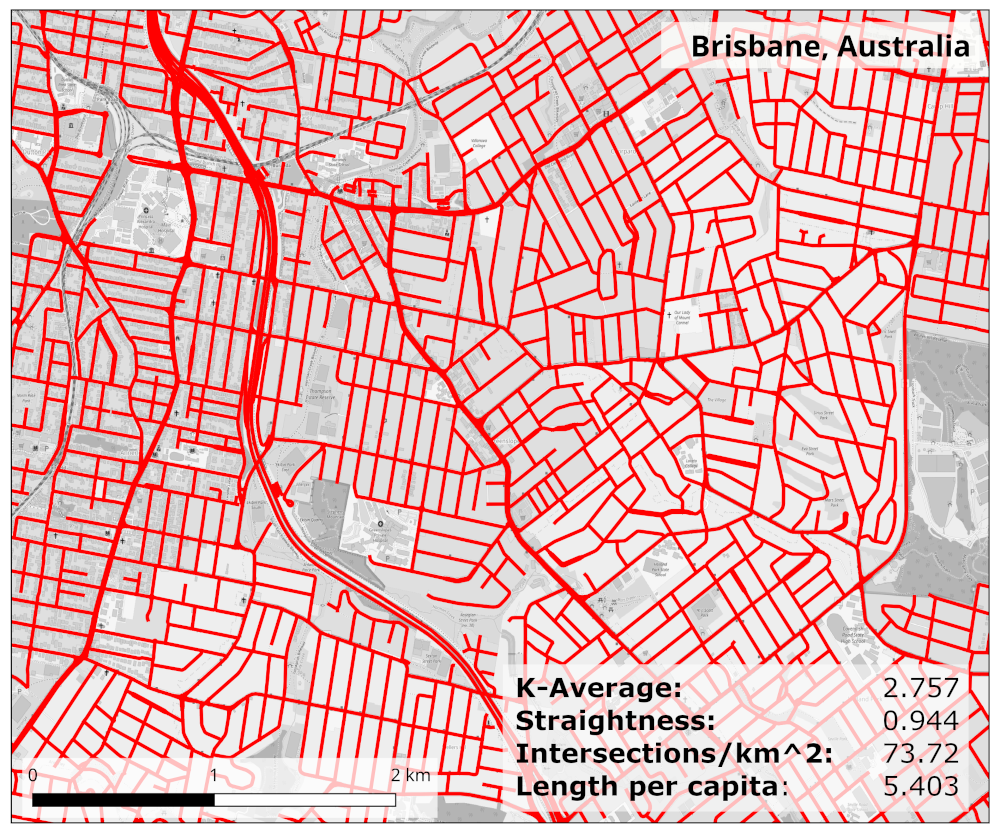}
        \caption{Low Density Deformed Grid} \label{fig:b}
    \end{subfigure}
    \medskip
    \begin{subfigure}{0.49\textwidth}
        \includegraphics[width=\linewidth]{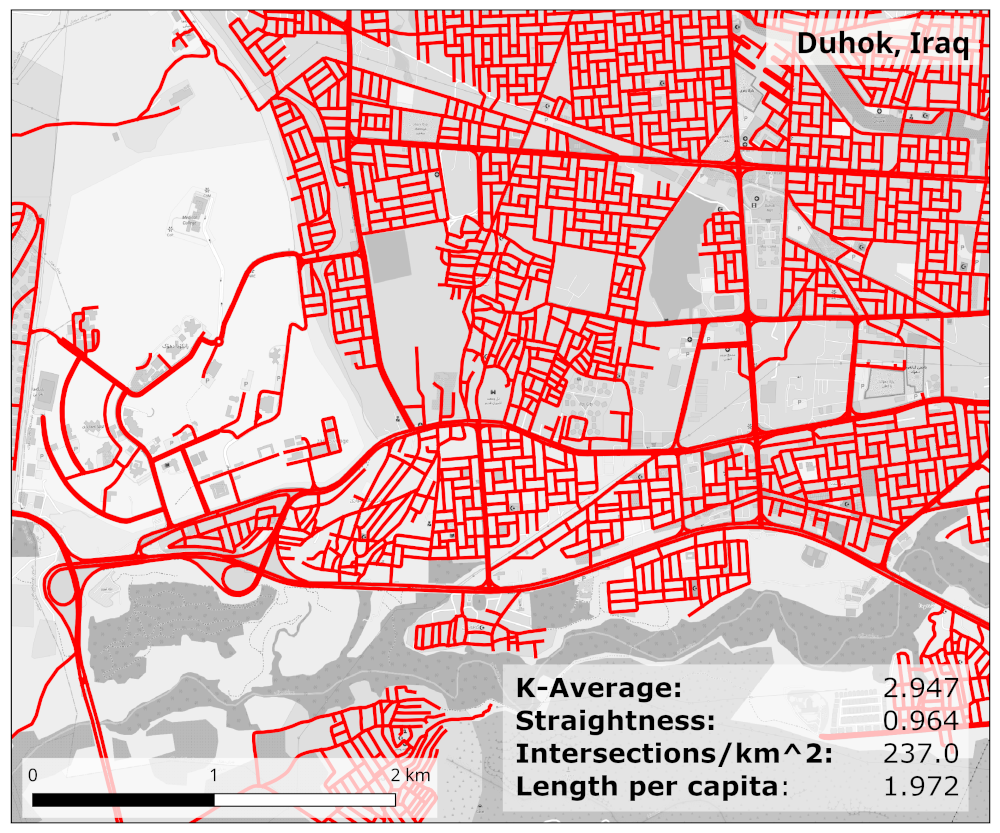}
        \caption{Medium Density Deformed Grid} \label{fig:c}
    \end{subfigure}\hspace*{\fill}
    \begin{subfigure}{0.49\textwidth}
        \includegraphics[width=\linewidth]{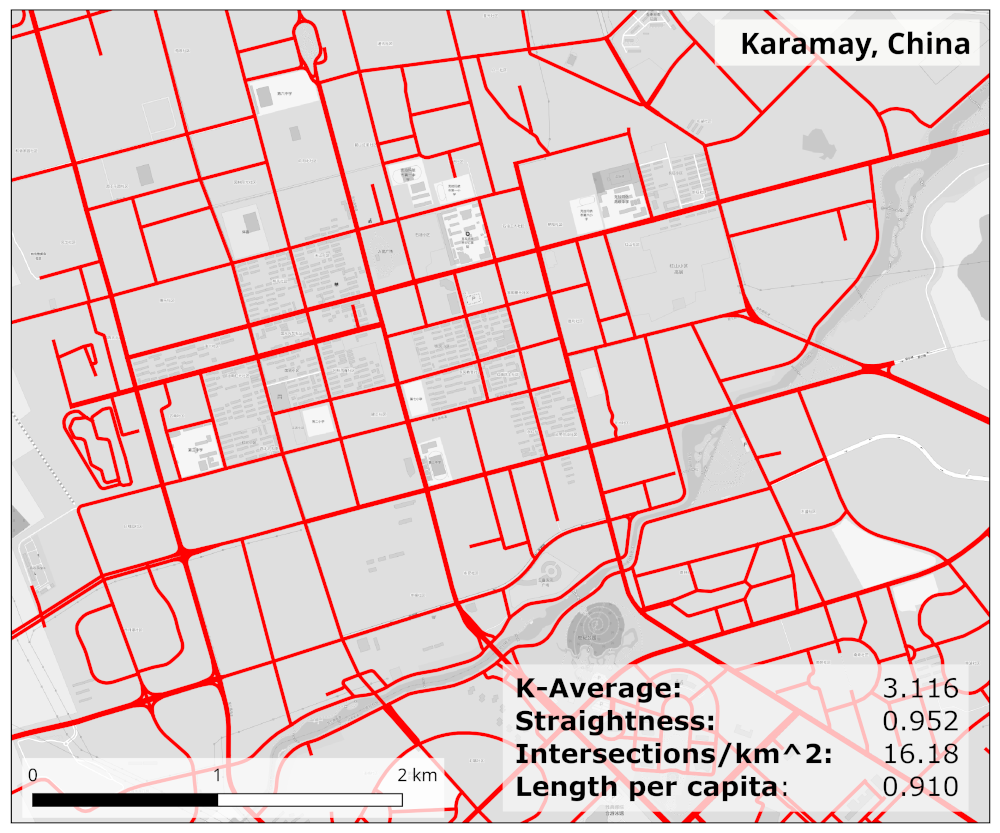}
        \caption{Modernist Superblocks} \label{fig:d}
    \end{subfigure}
    \medskip
    \begin{subfigure}{0.49\textwidth}
        \includegraphics[width=\linewidth]{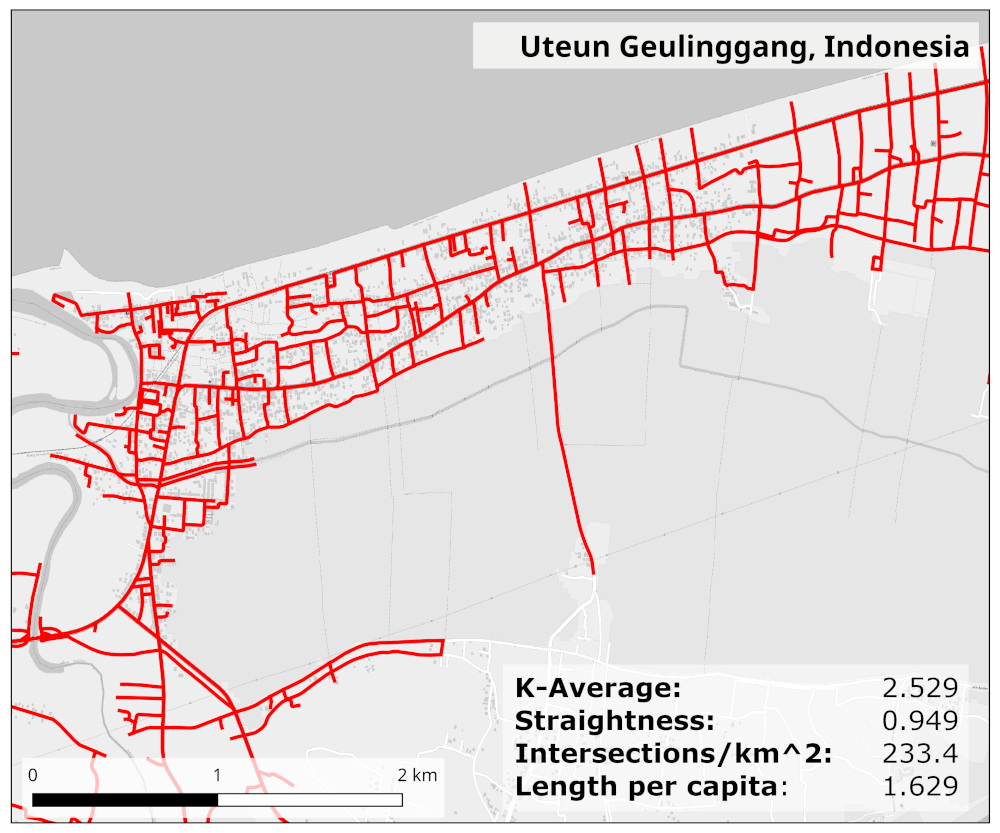}
        \caption{Dead Ends} \label{fig:e}
    \end{subfigure}\hspace*{\fill}
    \begin{subfigure}{0.49\textwidth}
        \includegraphics[width=\linewidth]{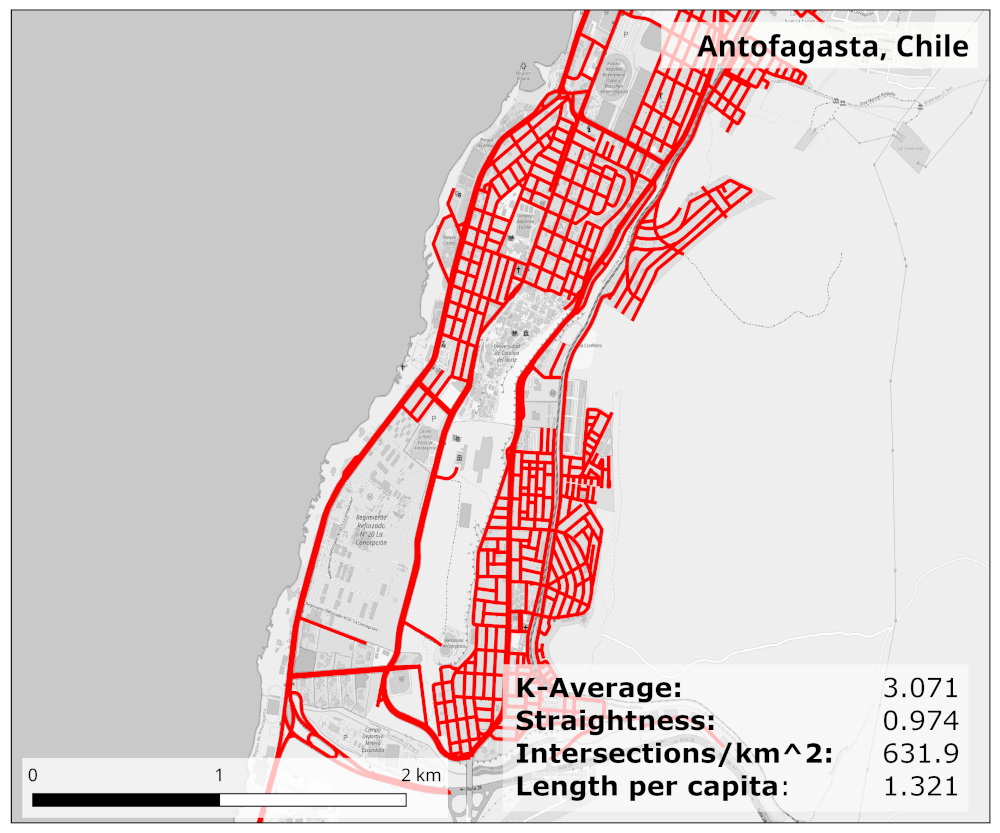}
        \caption{High Density Grid} \label{fig:f}
    \end{subfigure}
    \caption{Street networks exemplifying each of the six design paradigm clusters alongside four indicator values for that specific urban area.} \label{fig:urbanform_examples}
\end{figure}

Our cluster analysis reveals six clusters of urban areas that exhibit internally similar design optimizing within-group similarity and between-group dissimilarity. Figure \ref{fig:urbanform_examples} illustrates these clusters, which comprise: 1) modernist superblocks, most prevalent in China and post-Soviet countries, 2) low-density deformed grids, most prevalent in modern Western cities, 3) high-density networks dominated by dead-ends, most prevalent in India and other less-developed countries, 4) medium-density deformed grids, most prevalent in the Mediterranean and Latin America, 5) circuitous networks dominated by T-intersections, most prevalent in older cities in Western Europe and its former colonies, and 6) high-density grids, most prominent in less-developed countries.

\begin{figure}[htbp]
    \centering
    \includegraphics[width=1\linewidth]{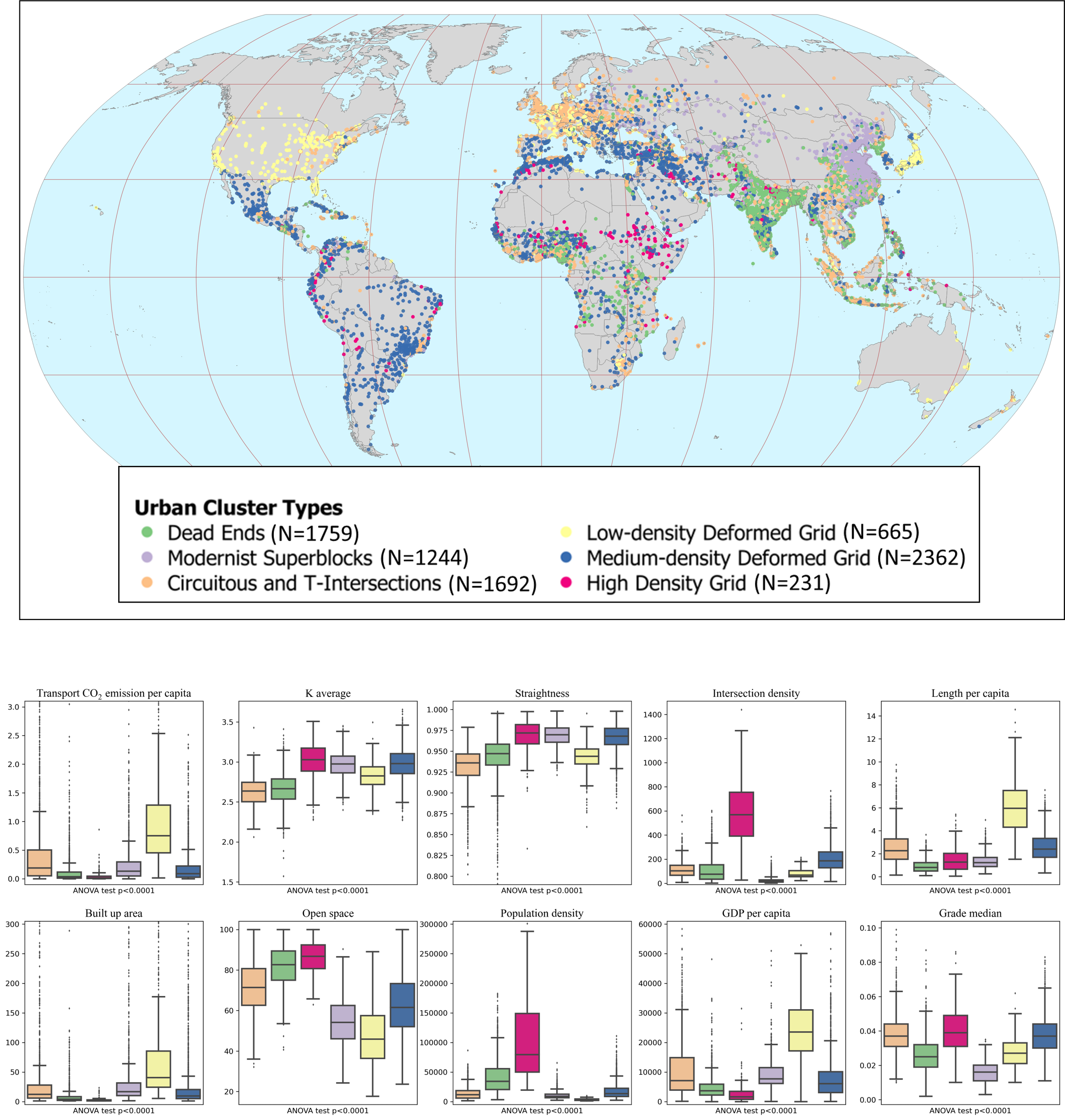}
    \caption{The spatial (above) and statistical (below) distributions of our six design paradigm clusters. See Table \ref{tab:variable_descriptions} for variable descriptions.}
    \label{fig:clusters_worldmap}
\end{figure}

Importantly, not all urban areas in a given region (or even country) belong to the design paradigm cluster most closely associated with that region, due to within-region historical divergences and design heterogeneity. Figure \ref{fig:clusters_worldmap} maps the world's urban areas by cluster and visualizes each cluster's statistical distributions across ten variables.

\begin{table}[htb]
\centering
\scriptsize
\caption{Model I's parameter estimates, standard errors (in parentheses), and standardized beta coefficients. Dummies not shown. Significance denoted as * $p<0.05$, ** $p<0.01$, *** $p<0.001$.}
\label{tab:regression_results}
\begin{tabular}{lrr}
	\toprule
	Variable & Parameter & Beta \\
	\midrule
	Intercept               &    1.025~~~  & \\
	                        &   (1.593)~~  & \\
	$k$ average (log)         &   -2.351***  & -0.088\\
	&   (0.269)  &\\
	Straightness              &    -5.245***     & -0.057 \\
	&   (0.811)  &\\
	Intersection density              &    1.735     & 0.011 \\
	&   (1.772)  &\\
	Length per capita (log)              &    0.762***     & 0.258 \\
	&   (0.059)  &\\
	Built up area (log)              &    1.426***     & 0.794 \\
	&   (0.035)  &\\
	Open space (log)             &    0.418***     & 0.050 \\
	&   (0.084)  &\\
	Population density (log)              &    0.945***     & 0.389 \\
	&   (0.062)  &\\
	GDP per capita (log)              &    0.068**     & 0.045 \\
	&   (0.026)  &\\
	Night light per capita (log)              &    0.123***     & 0.069 \\
	&   (0.021)  &\\
	Grade median              &    -13.258***     & -0.074 \\
	&   (1.537)  &\\
%	Airport              &    -0.048     & -0.007 \\
%	&   (0.049)  &\\
%	Waterport              &    -0.236     & -0.012 \\
%	&   (0.132)  &\\
	\midrule
	$n$                       &      7953       &        \\
	$R^2$               &      0.728       &      \\
	\bottomrule
\end{tabular}
\end{table}

\subsection{Regression Analysis}

Worldwide, in Model I (Table \ref{tab:regression_results}), we observe negative associations between street network connectedness and  transport-sector CO\textsubscript{2} emissions per capita, and between straighter streets and emissions. For example, a 1\% increase in connectedness (i.e., $k$ average)  is associated with a 2.4\% decrease in emissions. Furthermore, worldwide, we find positive relationships between per capita street length and emissions, but no significant relationship between intersection density and emissions. Even when controlling for population density, 1\% more street length per capita is associated with a 0.8\% increase in  emissions. Both measures of economic development in our model---night lights and GDP per capita---are associated with greater emissions.

However, these estimates represent average relationships aggregated across a wide range of cities differing substantially in urban form. Testing whether these relationships are the same in sign and similar in magnitude everywhere, our spatial regimes models estimate separate coefficients for each variable for each UN development group in Model II, and for each design paradigm cluster in Model III.

In Model II (Table \ref{tab:table3_new}), we observe heterogeneity in magnitude and statistical significance. While both connectedness and straightness are negatively associated with emissions everywhere, these relationships are only significant in LDCL urban areas---perhaps due to the larger sample size. The relationship between per capita street length and emissions is significant and positive across all three development groups, but is the largest in LDC urban areas, suggesting that the association between road (over-)building and emissions declines with rising development levels. However, these development levels themselves contain substantial within-group urban form heterogeneity.

Model III (Table \ref{tab:cluster_regression_results}) tests whether these relationships vary across design paradigms. Connectedness is consistently associated with lower emissions across all clusters, though the magnitude of this relationship is approximately three times as large in high density grid and modernist superblock urban areas than in the others. The relationship between straightness and  emissions---while not as consistent as connectedness---is also negative, save for the low density grid and dead ends clusters. Greater street lengths per capita on the other hand are associated with higher emissions in all clusters save for the high density grid cluster---that is, the urban areas that already have the shortest road lengths per capita. Finally, the relationship between intersection density and transport emissions is only statistically significant in the low density deformed grid cluster and in the modernist superblocks cluster---that is, the urban areas with the lowest intersection densities.

\begin{landscape}
	\begin{table}[htbp]
		\centering
        \footnotesize
		\caption{Model II's parameter estimates, standard errors (in parentheses), and standardized beta coefficients, across UN development groups. Dummies not shown. Significance denoted as * $p<0.05$, ** $p<0.01$, *** $p<0.001$.}
		\label{tab:table3_new}
		\begin{tabular}{lcccccc}
\toprule
      & \multicolumn{2}{c}{LDC} & \multicolumn{2}{c}{LDCL} & \multicolumn{2}{c}{MDR} \\
     \cmidrule(lr){2-3}\cmidrule(lr){4-5}\cmidrule(lr){6-7} &
     \multicolumn{1}{c}{parameter}  & \multicolumn{1}{c}{beta} & \multicolumn{1}{c}{parameter} & \multicolumn{1}{c}{beta} & \multicolumn{1}{c}{parameter} & \multicolumn{1}{c}{beta} \\
\midrule
 $k$ average (log) &  -1.296 &  -0.057   & -2.685*** &  -0.108   &  -0.624 & -0.030    \\
            & (0.965)  & & (0.334)  & & (0.369)  &  \\

 Straightness & -5.758 & -0.066 & -5.591*** &  -0.066 &  -1.881 &  -0.024  \\
  & (3.282) & & (1.008) & & (1.033) &  \\

 Intersection density &  -0.254 & -0.003 & 1.207 & 0.008 &  -16.556 &  -0.048  \\
 & (4.730) & & (2.177) & & (8.491) & \\

 Length per capita (log) &   1.102*** & 0.281 & 0.785*** & 0.260 & 0.313* & 0.107 \\
 & (0.198) & & (0.072) & & (0.123) & \\

 Built up area (log) &   1.573*** & 0.828 &  1.503*** & 0.811 &  1.039*** & 0.706  \\
    & (0.119) & & (0.045) & & (0.047) &  \\

 Open space (log) &  1.099** &  0.096  &  0.391** & 0.045 &  0.393*** & 0.078 \\
    & (0.424) & & (0.136) & & (0.058) & \\

 Population density (log) &  1.167*** & 0.372 &  1.034*** & 0.370 &   0.504*** & 0.169  \\
       & (0.207) & & (0.079) & & (0.127) &  \\

 GDP per capita (log) &   -0.052 & -0.034 &  0.056 &  0.035 &  0.188*** & 0.156  \\
   & (0.083) & & (0.033) & & (0.040) &  \\

 Night light per capita (log) &   0.162*** & 0.099 &  0.090*** &  0.051   & 0.098** & 0.043  \\
& (0.047) & & (0.028) & & (0.038) & \\

 Grade median & -6.683 & -0.037 & -15.291*** &  -0.095 & -2.786 & -0.018  \\
  & (5.562) & & (1.937) & & (2.010) &  \\
\midrule
    $n$ &  735 &  &  5708 & & 1565 &    \\
    $R^2$   & 0.711 &  & 0.629 & & 0.868 &  \\
\bottomrule
\end{tabular}
	\end{table}
\end{landscape}

\begin{landscape}
	\begin{table}[htbp]
		\centering
        \scriptsize
		\caption{Model III's parameter estimates, standard errors (in parentheses), and standardized beta coefficients, across design paradigm clusters. Dummies not shown. Significance denoted as * $p<0.05$, ** $p<0.01$, *** $p<0.001$.}
		\label{tab:cluster_regression_results}
		\begin{tabular}{lcccccccccccc}
\toprule
      & \multicolumn{2}{c}{Circuitous and T-intersections} & \multicolumn{2}{c}{Low Density Deformed Grid} & \multicolumn{2}{c}{Medium Density Deformed Grid} & \multicolumn{2}{c}{Modernist superblocks} & \multicolumn{2}{c}{Dead Ends} & \multicolumn{2}{c}{High density grid} \\
     \cmidrule(lr){2-3}\cmidrule(lr){4-5}\cmidrule(lr){6-7}\cmidrule(lr){8-9}\cmidrule(lr){10-11}\cmidrule(lr){12-13} &
     \multicolumn{1}{c}{parameter}  & \multicolumn{1}{c}{beta} & \multicolumn{1}{c}{parameter} & \multicolumn{1}{c}{beta} & \multicolumn{1}{c}{parameter} & \multicolumn{1}{c}{beta} & \multicolumn{1}{c}{parameter} & \multicolumn{1}{c}{beta} & \multicolumn{1}{c}{parameter} & \multicolumn{1}{c}{beta} & \multicolumn{1}{c}{parameter} & \multicolumn{1}{c}{beta} \\
\midrule
      $k$ average (log) &   -1.824** &  -0.054   & -1.500** &  -0.061   &  -1.596** & -0.047 & -4.663*** & -0.115 &  -1.761** & -0.056 &   -5.187* & -0.219    \\
            & (0.660)  & & (0.570)  & & (0.541)  & & (0.965)  & & (0.620  & & (2.085) & \\
  Straightness & -4.258* & -0.043 & -2.682 & -0.029  &  -13.963*** & -0.097 & -18.004*** & -0.105 & -0.377 & -0.005	& -16.598*  & -0.165 \\
  & (1.720)	& & (1.795) & & (2.228) & & (4.073) & & (1.545) & & (7.607) & \\
 Intersection density &  -8.676 & -0.025	& -102.619*** & -0.270	 & -2.331 & -0.012 & -133.809** & -0.173 &	-1.507 & -0.006 & 3.898 & 0.088   \\
 & (12.616) & & (20.849) & & (5.650) & & (48.483) & & (8.397) & & (4.535) & \\
 Length per capita (log) &   0.789*** & 0.203 & 1.204*** & 0.328 & 0.986*** & 0.224 & 1.445*** & 0.316 & 0.811*** & 0.222 & 0.519 & 0.231 \\
 & (0.224) & & (0.292) & & (0.174) & & (0.228) & & (0.148) & & (0.405) & \\
    Built up area (log) &   1.259*** & 0.654 & 0.945*** & 0.702 & 1.426*** & 0.769 & 1.487*** & 0.675 & 1.562*** & 0.700 & 1.755*** & 0.564 \\
    & (0.077) & & (0.115) & & (0.051) & & (0.133) & & (0.086) & & (0.252) & \\
    Open space (log) &  0.739** & 0.063 & 0.279*** & 0.061 & 0.416** & 0.046 & 0.202 & 0.021 & 1.718*** & 0.097 & 1.498 & 0.067 \\
    & (0.236) & & (0.077) & & (0.150) & & (0.311) & & (0.416) & (1.941) & \\
       Population density (log) &   0.844*** & 0.271 & 1.226*** & 0.393 & 1.134*** & 0.344 & 1.340*** & 0.321 & 1.078*** & 0.325 & 0.524 & 0.191 \\
       & (0.225) & & (0.301) & & (0.176) & & (0.262) & & (0.155) & & (0.512) & \\
   GDP per capita (log) &   0.130* & 0.086 & 0.254* & 0.218 & 0.100** & 0.071 & 0.116 & 0.066 &-0.071 & -0.040 & -0.006 & -0.004 \\
   & (0.059) & & (0.104) & & (0.036) & & (0.103) & & (0.063) & & (0.132) & \\
Night light per capita (log) &   0.204*** & 0.113 & 0.074 & 0.032 & 0.062 & 0.035 & 0.491*** & 0.202 & 0.014 & 0.008 & 0.064 & 0.052 \\
& (0.052) & & (0.070) & & (0.036) & & (0.076) & & (0.045) & & (0.094) & \\
  Grade median & -4.154 & -0.021 & 4.095 & 0.022 & -7.716** & -0.039 & -47.966*** & -0.141 & -17.644*** & -0.079 & -7.742 & -0.056 \\
  & (3.354) & & (3.430) & & (2.437) & & (8.473) & & (4.507) & & (8.745) & \\
\midrule
    $n$ &  1692 &  &  665 & & 2362 & &  1244 &  & 1759 &   & 231  &   \\
    $R^2$   & 0.741 &  & 0.870 & & 0.786 & & 0.534  &  & 0.598 &  & 0.666 &  \\
\bottomrule
\end{tabular}
	\end{table}
\end{landscape}

\begin{landscape}
	\begin{table}[htbp]
		\centering
        \scriptsize
		\caption{Summary of coefficients' signs and significance across regimes in Models II and III. Dummies not shown. Zeros indicate statistical insignificance at a 95\% confidence level.}
		\label{tab:cluster_regression_signs}
		\begin{tabular}{lccccccccc}
\toprule
                             &  &   &   & Circuitous and & Low Density & Medium Density & Modernist & & \\
                             & LDCL & LDC & MDR & T-intersections & Deformed Grid & Deformed Grid & Superblocks & Dead Ends &  High Density Grid \\
\midrule
$k$ average (log)              &  0	&	-	&	0	&	-	&	-	&	-	&	-	&	-	&	-        \\
Straightness                 &   0	&	-	&	0	&	0	&	0	&	-	&	-	&	0	&	-        \\
Intersection density         &   0	&	0	&	0	&	0	&	-	&	0	&	-	&	0	&	0        \\
Length per capita (log)      &  +	&	+	&	+	&	+	&	+	&	+	&	+	&	+	&	0        \\
Built up area (log)          &  +	&	+	&	+	&	+	&	+	&	+	&	+	&	+	&	+         \\
Open space (log)             &   +	&	+	&	+	&	+	&	+	&	+	&	0	&	+	&	0         \\
Population density (log)     &   +	&	+	&	+	&	+	&	+	&	+	&	+	&	+	&	0         \\
GDP per capita (log)         &   0	&	0	&	+	&	+	&	+	&	+	&	0	&	0	&	0          \\
Night light per capita (log) &  +	&	+	&	+	&	+	&	0	&	0	&	+	&	0	&	0          \\
Grade median                 &   0	&	-	&	0	&	0	&	0	&	-	&	-	&	-	&	0          \\
%\midrule
%$n$                          & 735  & 5708 & 1565 &              1692              &            665            &             2362             &         1244          &   1759    &        231         \\
\bottomrule
\end{tabular}

	\end{table}
\end{landscape}

Tables \ref{tab:regression_results}, \ref{tab:table3_new}, and \ref{tab:cluster_regression_results} also present standardized beta regression coefficients that estimate how many standard deviations of change in the response are associated with a one standard deviation change in a predictor. Table \ref{tab:cluster_regression_results} shows that even when the relationships between predictors and  transport-sector emissions possess the same sign, standardized magnitudes can vary dramatically between clusters. Increases in connectedness are associated with nearly twice as large a reduction in emissions in modernist superblock urban areas (which are common in China) than in low density deformed grid urban areas (which are common in North America). Table \ref{tab:cluster_regression_signs} summarizes the signs and significance of coefficient estimates across Models II and III.

\section{Discussion}

In recent years, a large body of literature has explored the relationships between street network design and transport emissions. Its goal---often complementing research on novel transport technologies or alternative fuels---is to inform planners and policymakers how urban form and street network design interventions could reduce GHG emissions and in turn mitigate global climate change. However, this literature's real world impact has been circumscribed by its regional or small sample research designs that have limited its applicability to other geographical contexts. This in turn has curtailed our knowledge of global relationships between transport planning and emissions. Yet climate change is a global phenomenon with global causes and global consequences. Planners around the world need a context-sensitive evidence base for their interventions into this crisis, and transport offers a key leverage point.

In this study we estimated these global relationships. Then, to unmask geographical heterogeneity, we re-estimated these relationships across development groups and design paradigm clusters. Our findings provide new universal estimates of the relationships between street network design and transport emissions while also providing insights into their heterogeneity. Worldwide, all else equal, we find that more-connected and straighter streets are associated with lower emissions, while greater street lengths per capita are associated with more emissions. This finding---that variables operationalizing denser urban form tend to be associated with lower transport-sector GHG emissions---is broadly consistent with current theory, but provides the first comprehensive global evidence of such in the literature.

At the same time, our models reveal important heterogeneity in these associations. Certain variables' relationships with transport emissions vary dramatically between different development levels or design paradigms. For instance, the association between greater intersection density and lower  transport-sector CO\textsubscript{2} emissions---present in our global model---appears to be limited to the least intersection-dense types of urban form, the low density deformed grid (that is, most cities in North America or Australia) and the modernist superblocks (that is, most cities in China and post-Soviet states). Similarly, longer road lengths per capita have a larger association with transport emissions in those same clusters, and no statistical relationship with emissions at all in the high density grid cluster.

One possible explanation for this pattern could be that the underlying relationships exhibit nonlinearity with diminishing returns at the margins. That is, at very low intersection densities, increasing intersection density and decreasing per capita road lengths may be associated with lower emissions, while this is not the case in places that already have high intersection densities \citep[cf.][]{cerin_determining_2022}. This nonlinearity may result from different types of cities---both in terms of income level and of network design---using transportation technologies differently, thus creating different relationships. For instance, gridded street networks with low intersection densities make automobile trips easier by design. They may also require greater detours for any given trip, compared to higher density grids.

Much of this literature's prior case study research was conducted in either the Global North or in China, both of which are outliers in terms of their relatively low intersection densities, as well as often exhibiting larger associations between street network characteristics and emissions. This calls into question how generalizable their findings are, and to what extent policy recommendations can be transferred to the rest of world.  Our finding of different relationships between clusters underscores how the literature's estimated relationships are inherently specific to their individual contexts. Although we find that the directions of the relationships are overall consistent with existing theory, our study raises implications for how practitioners should interpret the literature for evidence-informed planning. Since the relationships between urban form and transport emissions are heterogeneous and context-specific, applying a body of evidence from one well-studied region to a different less-studied region will often be inappropriate.

For example, if an urban planner in the Middle East---an under-studied geographical region in which most cities fall into the \enquote{medium density deformed grid} cluster---followed the global literature, they would overestimate the local relationship between greater intersection density and transport-sector CO\textsubscript{2} emissions. Our study provides this hypothetical planner a different---and more locally-calibrated---evidence base for their planning interventions to achieve local goals. Accordingly, practitioners and researchers must carefully consider whether their local context is sufficiently similar to any given study's context for the relationships to hold. Given the prevalence of such studies from the Global North, our findings illustrate the importance of developing a stronger empirical base for planning rapidly developing cities in the Global South.

This study also offers several opportunities for further research. First, our research design is cross-sectional and does not attempt to identify causal relationships. Identifying such causal relationships is an important next step for both scientific theorizing and policymaking. Further, the literature has yet to converge on an authoritative set of urban variables to include in model specifications explaining variations in transport emissions, or on an authoritative way to cluster urban areas into internally similar groups. This study---like any other---is unable to resolve this conundrum on its own. However, it taps into recent advances in the open science, open data, and open source movements to propose a set of variables that can be freely and consistently obtained for all urban areas worldwide in a uniform, well-documented manner. Refining these statistical controls is an important next step, as our variables and model specifications are neither perfect nor authoritative. More data gathering and local research---especially in under-studied regions---are essential to understand what variables should be included. In particular, equivalent data on land use entropy remains an important gap for such worldwide analyses and represents an important next step for research.

\section{Conclusion}

Climate change presents an urgent challenge to all planners. In particular, transport plays a critical role as it represents a significant source of GHG emissions. Sustainable urban planning requires more careful attention to designing and building transport infrastructure that reduces emissions and offers individuals a variety of mode choices beyond automobile dependence. However, planners need a clearer body of knowledge about what that infrastructure should look like. While the relationships between street network design and transport emissions are fairly well-studied in some regions, including the US, Europe, and China, they are under-studied in many of the most rapidly developing parts of the world. Practitioners often lack a strong evidence base for evidence-informed planning.

This study estimated relationships between street network design and per capita transport-sector CO\textsubscript{2} emissions across every urban area on the planet. We then tested whether these relationships vary between different levels of development or between different design paradigms to unpack global heterogeneity. We find that, all else equal, more-connected and straighter streets are associated with lower emissions, while greater street lengths per capita are associated with more emissions. However, our models reveal substantial heterogeneity in these relationships. We argue that evidence-informed practice derived from global models or dissimilar world regions could yield conclusions inapplicable to local planning contexts.

This paper unlocks future research in estimating these relationships worldwide and, in particular, at more local scales and in less-developed regions. As these important relationships vary between different kinds of urban areas, more evidence is needed for local planners to both set and meet climate and sustainability goals. Taking advantage of these kinds of data and models can build up that evidence base for a more sustainable transport future.

\section*{Acknowledgments}

This work was funded in part by a grant from the Pacific Southwest Region University Transportation Center and the U.S. Department of Transportation.

% print the footnotes as endnotes, if any exist
\IfFileExists{\jobname.ent}{\theendnotes}{}

% print the bibliography
\setlength{\bibsep}{0.00cm plus 0.05cm} % no space between items
\bibliographystyle{apalike}
\bibliography{references}

\begin{thebibliography}{}

\bibitem[Alfieri et~al., 2015]{hess-19-2247-2015}
Alfieri, L., Burek, P., Feyen, L., and Forzieri, G. (2015).
\newblock Global warming increases the frequency of river floods in europe.
\newblock {\em Hydrology and Earth System Sciences}, 19(5):2247--2260.

\bibitem[Barrington-Leigh and Millard-Ball, 2017]{barrington-leigh_worlds_2017}
Barrington-Leigh, C. and Millard-Ball, A. (2017).
\newblock The world's user-generated road map is more than 80\% complete.
\newblock {\em PLOS ONE}, 12(8):e0180698.

\bibitem[Barron et~al., 2014]{barron_comprehensive_2014}
Barron, C., Neis, P., and Zipf, A. (2014).
\newblock A {Comprehensive} {Framework} for {Intrinsic} {OpenStreetMap}
  {Quality} {Analysis}.
\newblock {\em Transactions in GIS}, 18(6):877--895.

\bibitem[Basiri et~al., 2016]{basiri_quality_2016}
Basiri, A., Jackson, M., Amirian, P., Pourabdollah, A., Sester, M., Winstanley,
  A., Moore, T., and Zhang, L. (2016).
\newblock Quality assessment of {OpenStreetMap} data using trajectory mining.
\newblock {\em Geospatial Information Science}, 19(1):56--68.

\bibitem[Boarnet and Crane, 2001]{Boarnet}
Boarnet, M. and Crane, R. (2001).
\newblock {\em Travel by design: The influence of urban form on travel}.
\newblock Oxford University Press.

\bibitem[Boarnet, 2011]{doi:10.1080/01944363.2011.593483}
Boarnet, M.~G. (2011).
\newblock A broader context for land use and travel behavior, and a research
  agenda.
\newblock {\em Journal of the American Planning Association}, 77(3):197--213.

\bibitem[Boeing, 2017]{BOEING2017126}
Boeing, G. (2017).
\newblock Osmnx: New methods for acquiring, constructing, analyzing, and
  visualizing complex street networks.
\newblock {\em Computers, Environment and Urban Systems}, 65:126--139.

\bibitem[Boeing, 2021]{boeing_off_2021}
Boeing, G. (2021).
\newblock Off the {Grid}…and {Back} {Again}?
\newblock {\em Journal of the American Planning Association}, 87(1):123--137.
\newblock Publisher: Routledge \_eprint:
  https://doi.org/10.1080/01944363.2020.1819382.

\bibitem[Boeing, 2022]{boeing_street_2021}
Boeing, G. (2022).
\newblock Street network models and indicators for every urban area in the
  world.
\newblock {\em Geographical Analysis}, 54(3):519--535.

\bibitem[Boeing et~al., 2022]{boeing_using_2022}
Boeing, G., Higgs, C., Liu, S., Giles-Corti, B., Sallis, J.~F., Cerin, E.,
  Lowe, M., Adlakha, D., Hinckson, E., Moudon, A.~V., Salvo, D., Adams, M.~A.,
  Barrozo, L.~V., Bozovic, T., Delclòs-Alió, X., Dygrýn, J., Ferguson, S.,
  Gebel, K., Ho, T.~P., Lai, P.-C., Martori, J.~C., Nitvimol, K., Queralt, A.,
  Roberts, J.~D., Sambo, G.~H., Schipperijn, J., Vale, D., Van~de Weghe, N.,
  Vich, G., and Arundel, J. (2022).
\newblock Using open data and open-source software to develop spatial
  indicators of urban design and transport features for achieving healthy and
  sustainable cities.
\newblock {\em The Lancet Global Health}, 10(6):e907--e918.

\bibitem[Boeing et~al., 2023]{doi:10.1177/00420980221145403}
Boeing, G., Lu, Y., and Pilgram, C. (2023).
\newblock Local inequities in the relative production of and exposure to
  vehicular air pollution in los angeles.
\newblock {\em Urban Studies}, page 00420980221145403.

\bibitem[Cao and Yang, 2017]{CAO2017480}
Cao, X. and Yang, W. (2017).
\newblock Examining the effects of the built environment and residential
  self-selection on commuting trips and the related co2 emissions: An empirical
  study in guangzhou, china.
\newblock {\em Transportation Research Part D: Transport and Environment},
  52:480--494.
\newblock Land use and transportation in China.

\bibitem[Cerin et~al., 2022]{cerin_determining_2022}
Cerin, E., Sallis, J.~F., Salvo, D., Hinckson, E., Conway, T.~L., Owen, N., van
  Dyck, D., Lowe, M., Higgs, C., Moudon, A.~V., Adams, M.~A., Cain, K.~L.,
  Christiansen, L.~B., Davey, R., Dygrýn, J., Frank, L.~D., Reis, R.,
  Sarmiento, O.~L., Adlakha, D., Boeing, G., Liu, S., and Giles-Corti, B.
  (2022).
\newblock Determining thresholds for spatial urban design and transport
  features that support walking to create healthy and sustainable cities:
  findings from the {IPEN} {Adult} study.
\newblock {\em The Lancet Global Health}, 10(6):e895--e906.

\bibitem[Cervero and Kockelman, 1997]{CERVERO1997199}
Cervero, R. and Kockelman, K. (1997).
\newblock Travel demand and the 3ds: Density, diversity, and design.
\newblock {\em Transportation Research Part D: Transport and Environment},
  2(3):199--219.

\bibitem[Corcoran et~al., 2013]{corcoran_analysing_2013}
Corcoran, P., Mooney, P., and Bertolotto, M. (2013).
\newblock Analysing the growth of {OpenStreetMap} networks.
\newblock {\em Spatial Statistics}, 3:21--32.

\bibitem[Crane, 2000]{doi:10.1177/08854120022092890}
Crane, R. (2000).
\newblock The influence of urban form on travel: An interpretive review.
\newblock {\em Journal of Planning Literature}, 15(1):3--23.

\bibitem[Curtis et~al., 1984]{doi:10.1061/(ASCE)0733-9488(1984)110:1(1)}
Curtis, F.~A., Neilsen, L., and Bjornsor, A. (1984).
\newblock Impact of residential street design on fuel consumption.
\newblock {\em Journal of Urban Planning and Development}, 110(1):1--8.

\bibitem[Dijkstra et~al., 2020]{dijkstra_how_2020}
Dijkstra, L., Hamilton, E., Lall, S., and Wahba, S. (2020).
\newblock How do we define cities, towns, and rural areas?

\bibitem[{European Environment Agency}, 2018]{eeg_2018}
{European Environment Agency} (2018).
\newblock Transport could burn up the eu’s entire carbon budget.
\newblock Technical report, European Environment Agency.

\bibitem[Ewing and Cervero, 2010]{doi:10.1080/01944361003766766}
Ewing, R. and Cervero, R. (2010).
\newblock Travel and the built environment.
\newblock {\em Journal of the American Planning Association}, 76(3):265--294.

\bibitem[Florczyk et~al., 2019]{florczyk_description_2019}
Florczyk, A., Melchiorri, M., Corban, C., Schiavina, M., Maffenini, L.,
  Pesaresi, M., Politis, P., Sabo, F., Carneiro, F. S.~M., Ehrlich, D., Kemper,
  T., Tommasi, P., Airaghi, D., and Zanchetta, L. (2019).
\newblock Description of the {GHS} {Urban} {Centre} {Database} 2015.
\newblock ISBN: 9789279997532.

\bibitem[Frenkel and Ashkenazi, 2008]{frenkel2008measuring}
Frenkel, A. and Ashkenazi, M. (2008).
\newblock Measuring urban sprawl: how can we deal with it?
\newblock {\em Environment and Planning B: Planning and Design}, 35(1):56--79.

\bibitem[Giles-Corti et~al., 2022a]{giles-corti_creating_2022}
Giles-Corti, B., Moudon, A.~V., Lowe, M., Adlakha, D., Cerin, E., Boeing, G.,
  Higgs, C., Arundel, J., Liu, S., Hinckson, E., Salvo, D., Adams, M.~A.,
  Badland, H., Florindo, A.~A., Gebel, K., Hunter, R.~F., Mitáš, J., Oyeyemi,
  A.~L., Puig-Ribera, A., Queralt, A., Santos, M.~P., Schipperijn, J.,
  Stevenson, M., Dyck, D.~V., Vich, G., and Sallis, J.~F. (2022a).
\newblock Creating healthy and sustainable cities: what gets measured, gets
  done.
\newblock {\em The Lancet Global Health}, 10(6):e782--e785.

\bibitem[Giles-Corti et~al., 2022b]{giles-corti_what_2022}
Giles-Corti, B., Moudon, A.~V., Lowe, M., Cerin, E., Boeing, G., Frumkin, H.,
  Salvo, D., Foster, S., Kleeman, A., Bekessy, S., de~Sá, T.~H.,
  Nieuwenhuijsen, M., Higgs, C., Hinckson, E., Adlakha, D., Arundel, J., Liu,
  S., Oyeyemi, A.~L., Nitvimol, K., and Sallis, J.~F. (2022b).
\newblock What next? {Expanding} our view of city planning and global health,
  and implementing and monitoring evidence-informed policy.
\newblock {\em The Lancet Global Health}, 10(6):e919--e926.

\bibitem[Grimm et~al., 2008]{doi:10.1126/science.1150195}
Grimm, N.~B., Faeth, S.~H., Golubiewski, N.~E., Redman, C.~L., Wu, J., Bai, X.,
  and Briggs, J.~M. (2008).
\newblock Global change and the ecology of cities.
\newblock {\em Science}, 319(5864):756--760.

\bibitem[Haklay, 2010]{haklay_how_2010}
Haklay, M. (2010).
\newblock How {Good} is {Volunteered} {Geographical} {Information}? {A}
  {Comparative} {Study} of {OpenStreetMap} and {Ordnance} {Survey} {Datasets}.
\newblock {\em Environment and Planning B: Planning and Design},
  37(4):682--703.

\bibitem[Handy et~al., 2002]{HANDY200264}
Handy, S.~L., Boarnet, M.~G., Ewing, R., and Killingsworth, R.~E. (2002).
\newblock How the built environment affects physical activity: Views from urban
  planning.
\newblock {\em American Journal of Preventive Medicine}, 23(2, Supplement
  1):64--73.
\newblock INNOVATIVE APPROACHES UNDERSTANDING AND INFLUENCING PHYSICAL
  ACTIVITY.

\bibitem[Hankey and Marshall, 2010]{HANKEY20104880}
Hankey, S. and Marshall, J.~D. (2010).
\newblock Impacts of urban form on future us passenger-vehicle greenhouse gas
  emissions.
\newblock {\em Energy Policy}, 38(9):4880--4887.
\newblock Special Section on Carbon Emissions and Carbon Management in Cities
  with Regular Papers.

\bibitem[Hong and Goodchild, 2014]{HONG201447}
Hong, J. and Goodchild, A. (2014).
\newblock Land use policies and transport emissions: Modeling the impact of
  trip speed, vehicle characteristics and residential location.
\newblock {\em Transportation Research Part D: Transport and Environment},
  26:47--51.

\bibitem[Jung et~al., 2022]{JUNG2022101223}
Jung, M.~C., Kang, M., and Kim, S. (2022).
\newblock Does polycentric development produce less transportation carbon
  emissions? evidence from urban form identified by night-time lights across us
  metropolitan areas.
\newblock {\em Urban Climate}, 44:101223.

\bibitem[Krey et~al., 2014]{krey_14_2014}
Krey, V., Masera, O., Blanford, G., Bruckner, T., Cooke, R., Fisher-Vanden, K.,
  Haberl, H., Hertwich, E., Kriegler, E., Mueller, D., Paltsev, S., Price, L.,
  Schlömer, S., Ürge Vorsatz, D., van Vuuren, D., and Zwickel, T. (2014).
\newblock 14: {Annex} {II}: {Metrics} \& {Methodolology}.
\newblock In {\em Climate {Change} 2014: {Mitigation} of {Climate} {Change}.
  {Contribution} of {Working} {Group} {III} to the {Fifth} {Assessment}
  {Report} of the {Intergovernmental} {Panel} on {Climate} {Change}}. Cambridge
  University Press, Cambridge, United Kingdom and New York, NY, USA.

\bibitem[Lashof and Ahuja, 1990]{Lashof}
Lashof, D.~A. and Ahuja, D.~R. (1990).
\newblock Relative global warming potentials of greenhouse gas emissions.
\newblock {\em Nature}, 344.6266:529--531.

\bibitem[Lemoine-Rodríguez et~al., 2020]{LEMOINERODRIGUEZ2020103949}
Lemoine-Rodríguez, R., Inostroza, L., and Zepp, H. (2020).
\newblock The global homogenization of urban form. an assessment of 194 cities
  across time.
\newblock {\em Landscape and Urban Planning}, 204:103949.

\bibitem[McPhearson et~al., 2016]{McPhearson2016}
McPhearson, T., Parnell, S., Simon, D., Gaffney, O., Elmqvist, T., Bai, X.,
  Roberts, D., and Revi, A. (2016).
\newblock Scientists must have a say in the future of cities.
\newblock {\em Nature}, 538(7624):165--166.

\bibitem[Mohajeri et~al., 2015]{MOHAJERI2015116}
Mohajeri, N., Gudmundsson, A., and French, J.~R. (2015).
\newblock Co2 emissions in relation to street-network configuration and city
  size.
\newblock {\em Transportation Research Part D: Transport and Environment},
  35:116--129.

\bibitem[Myhre et~al., 2019]{Myhre2019}
Myhre, G., Alterskj{\ae}r, K., Stjern, C.~W., Hodnebrog, {\O}., Marelle, L.,
  Samset, B.~H., Sillmann, J., Schaller, N., Fischer, E., Schulz, M., and
  Stohl, A. (2019).
\newblock Frequency of extreme precipitation increases extensively with event
  rareness under global warming.
\newblock {\em Scientific Reports}, 9(1):16063.

\bibitem[Reckien et~al., 2007]{doi:10.1080/00420980601136588}
Reckien, D., Ewald, M., Edenhofer, O., and Liideke, M. K.~B. (2007).
\newblock What parameters influence the spatial variations in co2 emissions
  from road traffic in berlin? implications for urban planning to reduce
  anthropogenic co2 emissions.
\newblock {\em Urban Studies}, 44(2):339--355.

\bibitem[Schweitzer and Zhou, 2010]{doi:10.1080/01944363.2010.486623}
Schweitzer, L. and Zhou, J. (2010).
\newblock Neighborhood air quality, respiratory health, and vulnerable
  populations in compact and sprawled regions.
\newblock {\em Journal of the American Planning Association}, 76(3):363--371.

\bibitem[Siew~Yin and Chin~Siong, 2010]{Siew_Yin_Chin_Siong_2010}
Siew~Yin, N. and Chin~Siong, H. (2010).
\newblock The relationship between urban population density and transportation
  fuel consumption in malaysian cities.
\newblock {\em Planning Malaysia}, 8(1).

\bibitem[Sims et~al., 2014]{sims_transport_2014}
Sims, R., Schaeffer, R., Creutzig, F., Cruz-Núñez, X., D’Agosto, M.,
  Dimitriu, D., Figueroa~Meza, M.~J., Fulton, L., Kobayashi, S., Lah, O.,
  McKinnon, A., Newman, P., Ouyang, M., Schauer, J.~J., Sperling, D., and
  Tiwari, G. (2014).
\newblock Transport.
\newblock In {\em Climate {Change} 2014: {Mitigation} of {Climate} {Change}.
  {Contribution} of {Working} {Group} {III} to the {Fifth} {Assessment}
  {Report} of the {Intergovernmental} {Panel} on {Climate} {Change}}. Cambridge
  University Press, Cambridge, United Kingdom and New York, NY, USA.

\bibitem[Taubenböck et~al., 2020]{TAUBENBOCK2020102814}
Taubenböck, H., Debray, H., Qiu, C., Schmitt, M., Wang, Y., and Zhu, X.
  (2020).
\newblock Seven city types representing morphologic configurations of cities
  across the globe.
\newblock {\em Cities}, 105:102814.

\bibitem[Uhl et~al., 2021]{uhl2021century}
Uhl, J.~H., Connor, D.~S., Leyk, S., and Braswell, A.~E. (2021).
\newblock A century of decoupling size and structure of urban spaces in the
  united states.
\newblock {\em Communications earth \& environment}, 2(1):20.

\bibitem[{US EPA}, 2021]{us_epa_2021}
{US EPA} (2021).
\newblock Fast {Facts}: {U.S.} {Transportation} {Sector} {Greenhouse} {Gas}
  {Emissions}, 1990-2019.
\newblock Technical report, United States Environmental Protection Agency.

\bibitem[Wang et~al., 2017]{WANG2017189}
Wang, S., Liu, X., Zhou, C., Hu, J., and Ou, J. (2017).
\newblock Examining the impacts of socioeconomic factors, urban form, and
  transportation networks on co2 emissions in china’s megacities.
\newblock {\em Applied Energy}, 185:189--200.

\bibitem[Xu et~al., 2018]{XU20181336}
Xu, L., Cui, S., Tang, J., Yan, X., Huang, W., and Lv, H. (2018).
\newblock Investigating the comparative roles of multi-source factors
  influencing urban residents' transportation greenhouse gas emissions.
\newblock {\em Science of The Total Environment}, 644:1336--1345.

\bibitem[Zielstra et~al., 2013]{zielstra_assessing_2013}
Zielstra, D., Hochmair, H.~H., and Neis, P. (2013).
\newblock Assessing the {Effect} of {Data} {Imports} on the {Completeness} of
  {OpenStreetMap} – {A} {United} {States} {Case} {Study}.
\newblock {\em Transactions in GIS}, 17(3):315--334.

\end{thebibliography}

\end{document}